\documentclass[
superscriptaddress,
amsmath,amssymb,
prl,
twocolumn,
]{revtex4-2}
\pdfoutput=1
\usepackage[utf8]{inputenc}
\usepackage[english]{babel}
\usepackage[T1]{fontenc}
\usepackage{amsmath}
\usepackage{tikz}
\usepackage{lipsum}
\usepackage{dsfont}
\usepackage[normalem]{ulem}
\usepackage{mathtools}
\usepackage[makeroom]{cancel}
\usepackage{bbm}

\usepackage[colorlinks,citecolor=blue]{hyperref}
\usepackage{graphicx}
\usepackage{amsthm}

\definecolor{ms}{rgb}{0,.4,1}

\newcommand{\tr}{\mathrm{Tr}} 
\newcommand{\Tr}[1]{\mathrm{Tr}\left[ #1\right]} 

\newcommand{\id}{\mathbbm{1}}

\newcommand{\RR}{\mathbb{R}}
\newcommand{\LL}{\mathbb{L}}

\newcommand{\ket}[1]{\left.\left|{#1}\right.\right\rangle}

\newcommand{\bra}[1]{\left.\left\langle{#1}\right.\right|}

\newcommand{\braket}[2]{\left\langle #1 \middle| #2 \right\rangle}

\newcommand{\ketbra}[2]{\ket{#1} \!\! \bra{#2}}

\newcommand{\sandwich}[3]
{\left\langle  #1 \right| #2 \left| #3 \right\rangle}

\newcommand{\average}[2]{\left \langle #1 \right \rangle_{#2}}
\newcommand{\rom}[1]{\uppercase\expandafter{\romannumeral #1\relax}}
\newcommand{\norbra}[1]{\left( #1\right)}
\newcommand{\norbraB}[1]{\boldsymbol{(}#1\boldsymbol{)}}
\newcommand{\sqrbra}[1]{\left[ #1\right]}

\usepackage{amssymb}             

\newcommand{\de}{{\rm d}}

\begin{document}
	
	\title{Universal Statistics of Charge Exchanges in Non-Abelian Quantum Transport}
	
	\date{\today}
	
	\begin{abstract}
		We derive detailed and intergral fluctuation relations as well as a Thermodynamic Uncertainty Relation constraining the exchange statistics of an arbitrary number of non-commuting conserved quantities among two quantum systems in transport setups arbitrary far from equilibrium. These universal relations, valid without the need of any efficacy parameter, extend the well-known heat exchange fluctuation theorems for energy and particle transport to the case of non-Abelian quantum transport, where the non-commutativity of the charges allows 
		{ going beyond standard thermodynamic behavior. In particular, we show that this can lead to 
			enhanced} precision in the current fluctuations, and it allows for the inversion of all currents against their affinity biases.
	\end{abstract}
	
	\author{Matteo Scandi}
	\affiliation{Instituto de F\'isica Te\'orica (IFT), UAM-CSIC, Madrid, Spain}
	\affiliation{Institute for Cross-Disciplinary Physics and Complex Systems (IFISC), UIB-CSIC, E-07122 Palma de Mallorca, Spain}
	
	\author{Gonzalo Manzano}
	\affiliation{Institute for Cross-Disciplinary Physics and Complex Systems (IFISC), UIB-CSIC, E-07122 Palma de Mallorca, Spain}
	
	\maketitle

	Non-commutativity is at the heart of quantum physics, underpinning fundamental concepts such as uncertainty, superposition, and entanglement. It enables most of its powerful and paradoxical properties: from the violation of Bell inequalities and Wigner's friend paradox~\cite{Bell04,Caslav22}, to transport effects such as negative differential conductance~\cite{rogge2006negative}, and quantum enhancements in computation and communication~\cite{Nielsen2000,Barlett2007}. With the advent of the emerging field of quantum thermodynamics~\cite{Deffner2019}, the interplay of conservation laws with non-commutativity has been realized as a key feature that can be associated with genuine quantum features at small scales~\cite{YungerHalpern16,Guryanova16,Lostaglio17}, and may allow possible thermodynamic enhancements in nonequilibrium situations~\cite{Manzano22}.
	
	Thermodynamics traditionally assumes that globally conserved quantities (or charges) in transport scenarios commute between them. Typical examples are the exchange of energy and particles due to temperature gradients or differences in chemical potentials, where the charges are observables represented by quantum operators that commute, for example $[H, N] = 0$. On the other hand, recent years have seen a raising interest in the thermodynamics of non-commuting charges (for a recent review see Ref.~\cite{Majidy23}), that is, situations where the observables associated to the globally conserved quantities fail to commute, e.g. different angular momentum or spin components, $[S_x, S_y] = iS_z$, as happens in certain types of spin chains that can be experimentally tested in the lab~\cite{Kranzl23}. Other systems showing non-commuting charges include bosonic systems in squeezed thermal states~\cite{Manzano18,Manzano16,Manzano22,Oliveira22,Shahidani24}, which has been used to implement nanomachines in the laboratory~\cite{Klaers17}.
	
	The absence of commutativity in the charges introduces nuances in the definition of thermal (equilibrium) states, where exact microcanonical ensembles cannot exist~\cite{YungerHalpern16,YungerHalpern20}, and their thermalization may deviate from the standard Eigenstate Thermalization Hypothesis~\cite{Majidy24}, leading to a non-commuting version of it~\cite{Murthy23}. Tradeoffs in the extraction and average exchange of charges that do not commute have also been considered from the point of view of resource theories~\cite{Guryanova16,Lostaglio17}, where separated charge-specific batteries may be used to store work~\cite{Popescu2020}. Reversible protocols have also been constructed for systems in the presence of baths equipped with non-commuting charges~\cite{Manzano18}. 
	
	More recently, the role of fluctuations in the transport of non-commuting charges has started to be considered~\cite{Manzano22}. In the linear regime, while the Onsager reciprocal relations~\cite{Onsager1,Onsager2} have been shown to hold, the fluctuation-dissipation relation~\cite{Callen,Kubo} is modified with the inclusion of an additional term due to the non-commutativity of the charges, which predicts reductions in the average entropy production with respect to the commutative case at the same level of fluctuations~\cite{Manzano22}. In addition, different definitions of entropy production that are equivalent for commuting charges can differ from each other~\cite{Upadhyaya24}, with fluctuations that may break standard universal relations from stochastic thermodynamics~\cite{Seifert12}. Similarly, non-commuting charges introduce subtleties in the thermodynamics of monitored quantum systems~\cite{Manzano22b}. 
	
	Arbitrarily far from equilibrium, classical and quantum systems obey the so-called Exchange Fluctuation Theorem (XFT)~\cite{Jarzynski04,Andrieux09} for a set of $N$ commuting charges $\{Q_1, Q_2, ..., Q_N\}$ with $[Q_i , Q_j] = 0~\forall i,j$, which reads:
	\begin{equation} \label{eq:XFT}
		\frac{P(\Delta \bf{Q})}{P(-\Delta \bf{Q})} =  e^{\sum_i \delta \lambda_i \Delta Q_i}, 
	\end{equation}
	where $\Delta {\bf{Q}} = \{\Delta Q_i \}$ represent stochastic exchanges in charges $\{{Q}_i\}$ during the process obtained from initial and final projective measurements, and $\delta \lambda_i$ are the corresponding differences in affinities associated with each charge (inverse temperature differences, chemical potential differences, etc.). The above equality is valid for (initially uncorrelated) systems starting in local equilibrium distributions~\cite{Campisi11,Esposito14}, 
	and has been experimentally tested on a variety of platforms~\cite{Nakamura10,Utsumi10,Gomez11,Hernandez20,Hernandez21}. Extensions have been derived which take into account initial classical correlations~\cite{Jevtic15}, while initial quantum correlations can be captured in a recent version of the XFT using quasiprobability distributions~\cite{Levy20}. Moreover, the XFT has been shown to be at the core of fundamental trade-off inequalities such as the Thermodynamic Uncertainty Relation~\cite{Hasegawa19,Timpanaro19}.
	
	However, the XFT \eqref{eq:XFT} and related equalities and inequalities are expected to break down in the case of non-commuting charges ($[Q_i,Q_j] \neq 0$ for some $i,j$), where the introduction of non-invasive measurement schemes to assess the fluctuations is also not obvious~\cite{Lostaglio23,Micadei20}. An interesting question is how non-commutativity precisely affects current fluctuations in arbitrary out-of-equilibrium regimes, and if new exchange fluctuation relations taking into account the non-commutativity effects may arise in this case. In this Letter, we answer this question in the affirmative by deriving exchange fluctuation theorems for non-Abelian transport [Eqs.~\eqref{eq:fluctuationsAverage} and \eqref{eq:Jarzynski}]. These universal relations include an extra term accounting for the non-commutativity of the charges and provide a more accurate version of the second law governing the tradeoffs between currents [Eq.~\eqref{eq:secondlaw}] and of the Thermodynamic Uncertainty Relation [Eq.~\eqref{eq:FTUR}]. 
	
	
	{\it Framework.} A generalized Gibbs state is uniquely identified by a set of scalars ${\bf \lambda} = \{\lambda_i\}$, called the affinities, and a set of associated observables ${\bf Q} = \{Q_i\}$ or charges, corresponding to globally conserved quantities. 
	Then, the generalized Gibbs state is given by~\cite{Jaynes57,Jaynes57b}:
	\begin{align}
		\pi_{\bf \lambda} = \frac{e^{-\sum_i\, \lambda_i Q_i}}{Z} = \frac{e^{-\mathcal{H}}}{Z}
	\end{align}
	where $Z= \tr[e^{-\sum_i \lambda_i Q_i}]$ is the partition function, and we have introduced the operator $\mathcal{H}:=\sum_i\, \lambda_i Q_i$. 
	Generalized Gibbs states have traditionally been explored in the context of integrable and near-integrable systems~\cite{Rigol07,Rigol09,Langen15} and for commuting charges~\cite{Vaccaro11,Vaccaro17,Mur18}. For situations where the charges do not commute with each other, $([Q_i, Q_j] \neq 0)$, as we consider here, they have also been called non-Abelian thermal states~\cite{YungerHalpern16,Majidy23}.
	
	We consider the following setting: two baths $A$ and $B$ {\color{black} with local Hamiltonians $H_A$ and $H_B$,} have been separately equilibrated to  generalized Gibbs states $\pi_{\bf \lambda^{A}}^{A}$ and $\pi_{\bf \lambda^{B}}^{B}$ each with their own set of conserved charges, $\{Q_i^{A} \}$ and $\{Q_i^{B} \}$, each representing the same globally conserved quantity, but different affinities ${\bf \lambda^A}$ and ${\bf \lambda^B}$.
	Our goal is to study the currents of (globally conserved) non-commuting charges that are generated when the two baths are put in contact with each other. In order to do so, we consider a collisional model for these exchanges~\cite{Manzano22} which can be divided into three steps:
	\begin{enumerate}
		\item An environmental unit (molecule) is taken from each bath, leading to an initial product state $\rho_0 = \pi^A_{\lambda^A} \otimes \pi^B_{\lambda^B}$, and are measured by projectors $\Pi^A$ and $\Pi^B$ in the eigenbasis of $\mathcal{H}_{A}$ and $\mathcal{H}_{B}$ respectively~\footnote{{ By choosing the measurement projectors proportional to operators $\mathcal{H}_X = \sum_i \lambda_i Q_i^X$ for $X=A,B$ we ensure that the initial generalized Gibbs state, $\rho_0$, is not altered by the measurement backaction, since $[\mathcal{H}_X, \pi_{\lambda^X}^X] = 0$.}};
		\item The two units interact through the unitary operation $U = \mathbb{T} e^{-i \int_0^\tau dt H_{\rm int}(t)}$ ({\color{black} in the interaction picture}), where $H_{\rm int}$ is a (possibly time-dependent) interaction Hamiltonian that couples the two baths during the interaction time $\tau$, and $\mathbb{T}$ is the time-ordering operator;
		\item The units are measured again in the eigenbasis of $\mathcal{H}_{A}$ and $\mathcal{H}_{B}$ respectively, after which they return to their respective baths. Then a new collision takes place repeating this procedure.
	\end{enumerate}
	In order for the currents to be considered as arising purely from the gradient in affinities between the two baths, one has to enforce the following condition for the interaction between the bath units for all charges $i$:
	\begin{align}
		\sqrbra{U, Q^A_i + Q^B_i} = 0\,,\label{eq:commutation}
	\end{align}
	which corresponds to the requirement that the unitary does not create extra charges, i.e., that the interaction is charge preserving (i.e., $[H_{\rm int}(t),\, Q^A_i + Q^B_i]=0~~\forall t \in [0,\tau]$). {\color{black} Eq.~\eqref{eq:commutation} corresponds to requiring that the interaction does not affect the charges at a global level. Moreover, we also assume that the charges cannot change locally (in the absence of interaction), $[H_A + H_B, Q^A_i + Q_i^B] = 0$ for all $i$. In other words, each charge is globally conserved during the evolution~\footnote{{\color{black} Strictly speaking, this extra condition is not necessary to derive our results, but it is needed in order to interpret $Q^A_i$ and $Q_i^B$ as conserved charges \emph{before} the onset of the interaction.}}}. We notice that this constraint is the only one that sets apart the framework studied here from the usual thermal setting with a single charge corresponding to the local Hamiltonians $Q_0^A= H_A$ and $Q_0^B = H_B$ and the affinities corresponding to the inverse temperatures. Indeed, in that case the only requirement is that $U$ commutes with the local Hamiltonians, $[U, H_A+H_B] = 0$, meaning that no work is introduced in the systems when the two baths interact, a condition that is typically known as strict energy conservation~\cite{Brandao13,Horodecki13,Korzekwa16}. In the general case, the extra set of constraints in Eq.~\eqref{eq:commutation} restricts the allowed interactions and conceptually justifies the interpretation of transport for the exchange of charges.
	
	
	\begin{figure}
		\centering
		\includegraphics[width=0.9\linewidth]{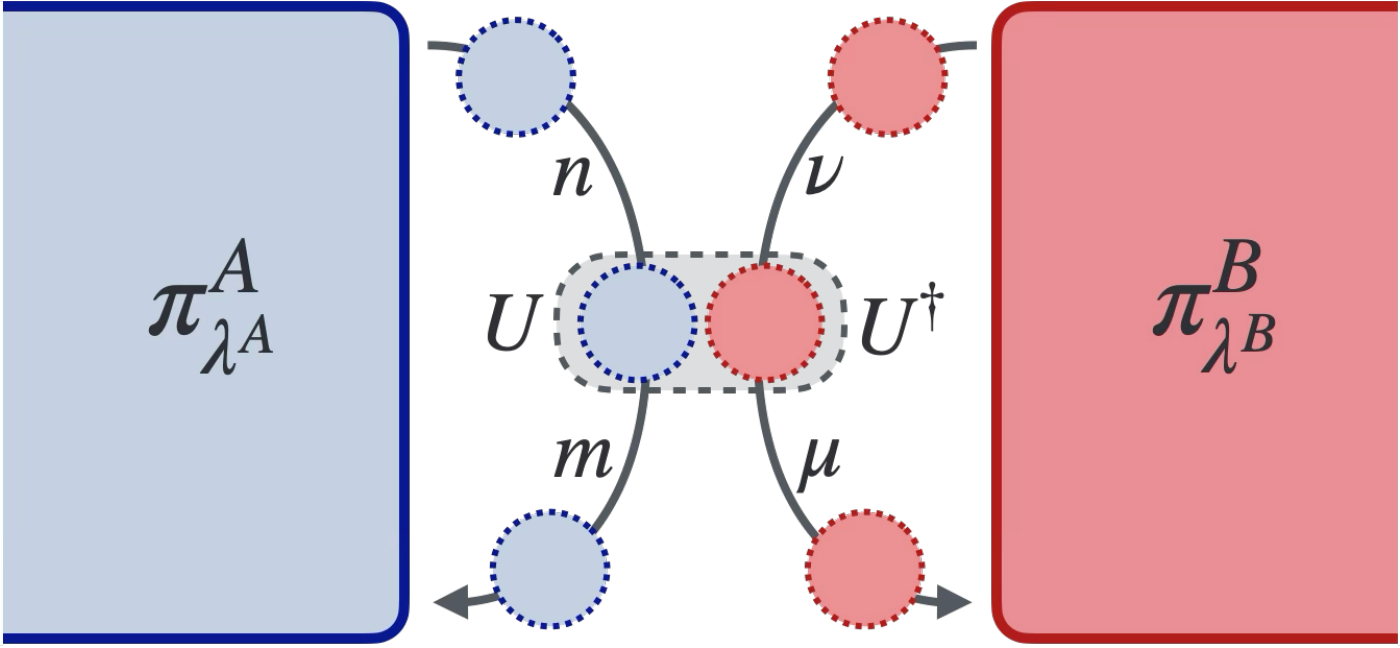}
		\caption{Depiction 
			of the collisional exchange model. Particles from each bath are projectively measured in the eigenbasis of $\mathcal{H}_A$ ($\mathcal{H}_B$) giving outcome $n$ ($\nu$), they interact with each other through the charge-preserving unitary operation $U$, and, finally, they are measured again giving outcome $m$ ($\mu$) before being returned to their original baths.}
		\label{fig:framework}
	\end{figure}
	
	Given this situation, it is interesting to explore how Eq.~\eqref{eq:commutation} affects the statistics of charge exchanges in the collisions. To this end, let us introduce the notion of a trajectory $\gamma := \{(n,\nu), (m,\mu)\}$ from the projective measurement outcomes in the process. This corresponds to the case in which an experimenter performs the first measurements in both units in the eigenbasis of $\mathcal{H}_{A}$ and $\mathcal{H}_{B}$ obtaining results $h_n^A$ and $h_\nu^B$ associated with outputs $n$ and $\nu$ (greek letters are reserved for units of bath $B$). Then the units evolve according to $U$. For simplicity, we assume that the interaction Hamiltonian $H_{\rm int}$ is suddenly switched on and off at the beginning and at the end of the interval, i.e., $H_{\rm int}(0) = H_{\rm int}(\tau) = 0$, and that it is constant for $0<t<\tau$. The experimenter finally performs a new measurement of $\mathcal{H}_A$ and $\mathcal{H}_B$ obtaining results $h_m^A$ and $h_\mu^B$ associated with outputs $m$ and $\mu$ (see Fig.~\ref{fig:framework}). In the following, we will use the eigendecomposition of $\mathcal{H}_A:= \sum_i \, h_i^A\,\ketbra{h^A_{i}}{h^A_{i}}$, together with the notation $\Delta h^{A}(n\rightarrow m): =  h^A_m - h^A_n$ (and similarly for bath $B$). Finally, it is also convenient to rewrite the affinities of bath $A$ as $\lambda^{A}_i = \lambda^B_i+\delta\lambda_i$. 
	
	Using the above framework, it is easy to compute the probability associated to the trajectory $\gamma$ representing a single collision, which reads:
	\begin{align}
		\mathbb{P}(\gamma) &= \frac{e^{-(h_n^A+h_\nu^B)}}{Z}\;|\!\sandwich{h_{m,\mu}}{U}{h_{n,\nu}}|^2\,,\label{eq:pathProbability}
	\end{align}
	where $\ket{h_{n,\nu}}$ are product state eigenvectors $\ket{h_{n,\nu}}\equiv\ket{h^A_{n}}\otimes\ket{h^B_{\nu}}$ (and similarly for $\ket{h_{m,\mu}}$. Then, as detailed in the Supplemental Material (see Sec.~\ref{app:derSelectionRule}), we derive the conservation relation:
	\begin{equation}
		\begin{split}
			\Delta h^A({n\rightarrow m}) &+ \Delta h^B({\nu\rightarrow \mu}) =\\&= \sum_{i} \;\delta\lambda_i\, \Delta Q_i({n\rightarrow m}) + \Delta({\gamma})\,,
		\end{split}\label{eq:selectionRuleCompact}
	\end{equation}
	where $\Delta Q_i ({n\rightarrow m}):= \sandwich{h_m^A}{Q^A_i}{h_m^A}-\sandwich{h_n^A}{Q_i^A}{h_n^A}$, that is the change in the expectation value of charge $Q^A_i$ when passing from $\ket{h_n^A}$ to $\ket{h_m^A}$, whereas $\Delta(\gamma)$ is a stochastic correction term reading:
	\begin{align}
		&\Delta(\gamma) = \sum_{\substack{i,k\\k\neq m}} \;\delta\lambda_i \norbra{\bra{h_m^A}Q_i^A\ket{h_{k}^A}\frac{\bra{h_{k,\mu}}H_{\rm int} \ket{h_{n,\nu}}}{\bra{h_{m,\mu}} H_{\rm int}\ket{h_{n,\nu}}}} +\nonumber\\
		&-\sum_{\substack{i,k\\k\neq n}}\delta\lambda_i\norbra{\bra{h_k^A}Q_i^A\ket{h_{n}^A}\frac{\bra{h_{m,\mu}}H_{\rm int} \ket{h_{k,\nu}}}{\bra{h_{m,\mu}} H_{\rm int} \ket{h_{n,\nu}}}}\,.\label{eq:DeltaExpr}
	\end{align}
	Two properties of $\Delta(\gamma)$ allow us to clarify its meaning (see the Supplemental Material Sec.~\ref{app:derSelectionRule} for the proofs): first, $\Delta(\gamma)$ is a real number, and it is zero for all trajectories if all charges commute with each other (that is, $[Q_i, Q_j] = 0$ for all $i,j$). For this reason, $\Delta(\gamma)$ can be identified as a signature of the lack of commutativity of charges. Secondly, when defining an analogous correction $\Delta({\tilde\gamma})$ associated with the reversed trajectory $\tilde\gamma: = \{(m,\mu), (n,\nu)\}$ coming from the inverse transition $\ket{h_{m, \mu}} \rightarrow \ket{h_{n, \nu}}$, we have that $\Delta({\tilde\gamma}) = -\Delta(\gamma)$. These two properties are used in the following to derive non-trivial fluctuation relations for non-commuting charges.
	
	{\it Non-Abelian Fluctuation Relations.} We now proceed to derive universal exchange fluctuation relations at the level of trajectories in the framework discussed above. These connect the probability associated with the trajectory $\gamma := \{(n,\nu), (m,\mu)\}$ with the probability of its time-reversed trajectory $\tilde\gamma:= \{(m,\mu), (n,\nu)\}$ in the backward process. A formal definition of the backward (or reversed) process can be found in the Supplemental Material, Sec.~\ref{app:derTimeReversal}. We obtain
	\begin{align} \label{eq:trajprobs}
		\frac{\mathbb{P}(\gamma)}{\mathbb{P}(\tilde{\gamma})} &= \frac{e^{-(h_n^A+h_\nu^B)}}{e^{-(h_m^A+h_\mu^B)}}\;\frac{|\!\sandwich{h_{m,\mu}}{U}{h_{n,\nu}}|^2}{|\!\sandwich{h_{n,\nu}}{U^\dagger}{h_{m,\mu}}|^2} \nonumber \\&= e^{\Delta h^A({n\rightarrow m}) + \Delta h^B({\nu\rightarrow \mu}) } \,,
	\end{align}
	which is valid for generic trajectories $\gamma$ with $\mathbb{P}(\tilde{\gamma}) \neq 0$ whenever $\mathbb{P}(\gamma) \neq 0$, that is, whenever the system does not present absolute irreversibility~\cite{Murashita14}. This relation follows from micro-reversibility of the unitary evolution $\tilde{U} = \Theta U^\dagger \Theta^{-1}$ with $\Theta$ the anti-unitary time-reversal operator~\cite{Campisi11,Manzano22b}{, and allow us to interpret $S_{\rm tot}(\gamma) \equiv \ln[\mathbb{P}(\gamma)/\mathbb{P}(\tilde{\gamma})]= \Delta h^A + \Delta h^B$ as the stochastic (information-theoretical) entropy production (EP) accounting for the irreversibility of the process~\cite{Manzano18b}}. For simplicity, we assumed that the interaction Hamiltonian and the charges are time-reversal symmetric,
	i.e., they do not contain odd variables under time reversal such as velocities or momenta. That assumption makes the backward process essentially equal to the forward one, leading to same probability measures in the numerator and denominator in the l.h.s. of Eq.~\eqref{eq:trajprobs}. Nonetheless, this assumption can be lifted, and we do so in Supplemental Material, Sec.~\ref{app:derTimeReversal}. 
	Combining Eq.~\eqref{eq:trajprobs} with Eq.~\eqref{eq:selectionRuleCompact} we obtain the relation:
	\begin{align}
		\frac{\mathbb{P}(\gamma)}{\mathbb{P}(\tilde{\gamma})} = e^{\sum_{i} \delta\lambda_i \Delta Q_i({n\rightarrow m})\, +\, \Delta({\gamma}) }\,.\label{eq:fluctuationTrajectories}
	\end{align}
	This is valid at the trajectory level but, thanks to its functional dependency, it allows us to group together trajectories with the same average exchanges of charges. In particular, we can define the { (bona-fide)} probability distribution for charge currents as:
	\begin{align}
		&P(\Delta Q_1,\dots, \Delta Q_N; \Delta) =\sum_{\gamma} \; \mathbb{P}(\gamma) \, \delta\norbra{\Delta Q_1- \Delta Q_1({n\rightarrow m})} \nonumber \\ & \times \dots \delta\norbra{\Delta Q_N- \Delta  Q_N({n\rightarrow m})}~\delta\norbra{\Delta - \Delta(\gamma)}\,.
	\end{align}
	Then, grouping together trajectories with the same $\Delta Q_i$ and $\Delta$ allows us to derive our main result, namely, the XFT for non-commuting charges:
	\begin{align}
		\frac{P(\Delta Q_1,\dots, \Delta Q_N; \Delta)}{P(-\Delta Q_1,\dots, -\Delta Q_N; - \Delta)} =  e^{\sum_{i} {\delta\lambda_i \Delta Q_i}\,+\,\Delta}\,,\label{eq:fluctuationsAverage}
	\end{align}
	which puts stringent constraints on the tails of the current probability distribution of the charges ${Q_i}$ (the proof and a generalization for nonsymmetric charges and interaction are provided in the Supplemental Material, Sec.~\ref{app:derFluctuations}). The above XFT generalizes the standard XFT in Eq.~\eqref{eq:XFT} derived in Refs.~\cite{Jarzynski04,Andrieux09}. In this context, it should be noticed that the presence of the quantum correction $\Delta$ is fundamental for the validity of the non-Abelian XFT above: indeed, trajectories $\gamma$ with the same change in the charges $\Delta Q_i(n \rightarrow m)$, can still differ in their value of $\Delta(\gamma)$. 
	
	The above detailed fluctuation relation implies the following integral (Jarzynski-like) version of the XFT:
	\begin{align}
		\average{e^{-\sum_{i} {\delta\lambda_i \Delta Q_i}\, -\, \Delta}}~ =  1\,,\label{eq:Jarzynski}
	\end{align}
	where the averages are taken with respect to the probability distribution $P(\Delta Q_1,\dots, \Delta Q_N; \Delta)$. As shown in the Supplemental Material Sec.~\ref{app:derStatistical}, the presence of the quantum correction $\Delta$ in the integral XFT above allows for a probability distribution of $\sum_{i} {\delta\lambda_i \Delta Q_i}$ that is more skewed towards negative values compared to the case of commuting charges.
	
	\begin{figure*}[th]
		\centering
		\includegraphics[width=.92\linewidth]{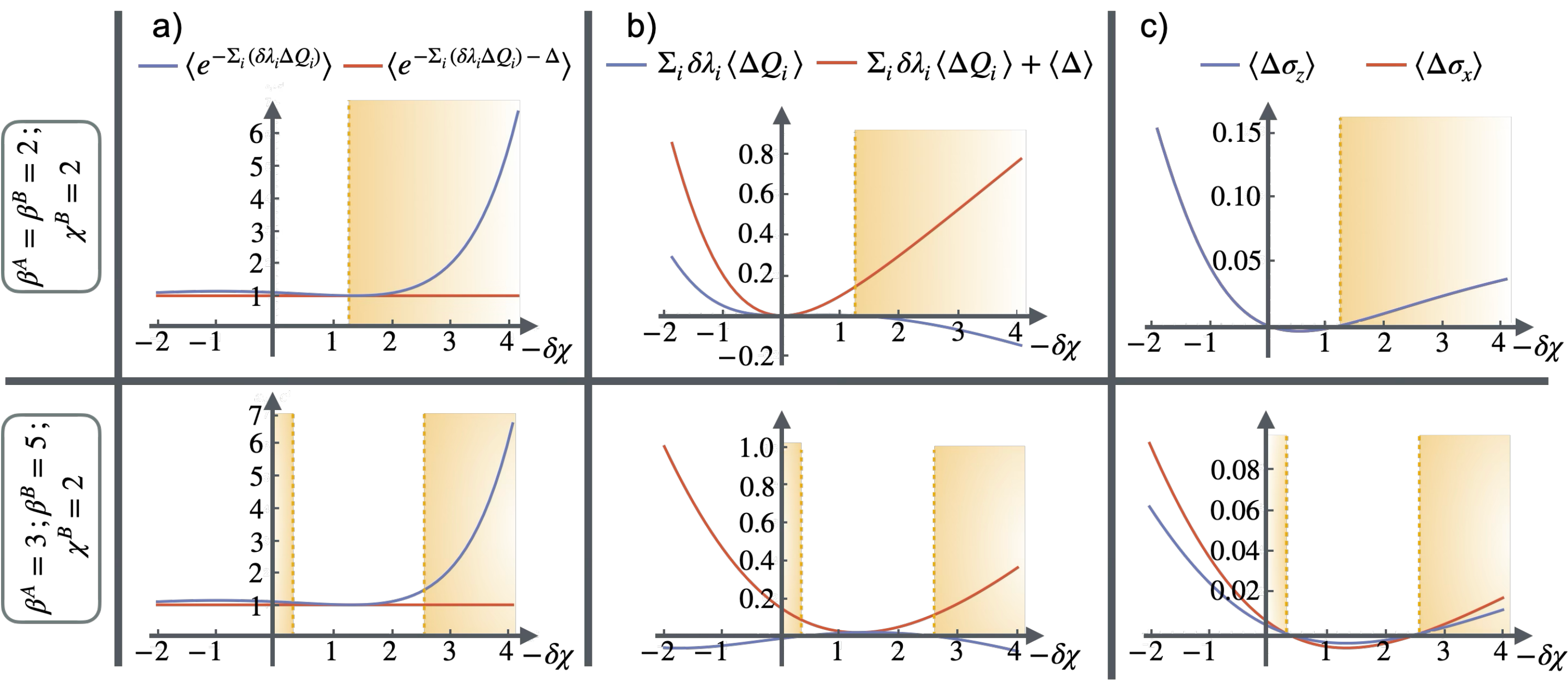}
		\caption{
			a) Integral XFT in Eq.~\eqref{eq:Jarzynski}, together with its standard version for commutative charges; b) second-law inequality in Eq.~\eqref{eq:secondlaw} with and without the quantum correction; c) average currents (in the top panel $\average{\Delta\sigma_{z}}{}\equiv\average{\Delta\sigma_{x}}{}$). Top row corresponds to same affinity $\beta$ in the baths ($\beta^A = \beta^B$), whereas in the bottom row $\beta^A<\beta^B$. Shaded areas corresponds to double current inversions. In the bottom row, for the parameter range between the two orange lines the second law holds even without quantum correction. This is reflected in a more tame behavior of the standard integral XFT.}
		\label{fig:simulations}
	\end{figure*}
	
	Applying Jensen's inequality, Eq.~\eqref{eq:Jarzynski} automatically implies the second-law inequality:
	\begin{align}
		\sum_{i} \delta\lambda_i\average{ \Delta Q_i}~\geq -\average{\Delta},~\label{eq:secondlaw}
	\end{align}
	where the average currents (charge changes per collision) are { $\average{ \Delta Q_i}{}=\tr[Q_i^A (\rho_1 - \rho_0)]$ with $\rho_1 = \sum_{m,\mu} (\Pi_m^A \otimes \Pi_\mu^B) U \rho_0 U^\dagger (\Pi_m^A \otimes \Pi_\mu^B)$ the evolved state after the second measurement, and the correction term $\average{\Delta}{}=
		\average{S_{\rm tot}}{}- \sum_{i} \;\delta\lambda_i\, \average{ \Delta Q_i}{}$, which 
		represents the difference between the average (information-theoretical) entropy production in a collision and its expected value from the measured currents.
		We remark that $\average{\Delta}{}$ can be either positive (currents apparently decrease EP) or negative (currents apparently increase EP), and becomes identically zero if 
		all charges commute with each other.} 
	Then the presence of the non-Abelian quantum correction $\langle \Delta \rangle$ in Eq.~\eqref{eq:secondlaw} 
	allows for { anomalous behavior in the average charge currents in apparent violation of the second law} (when $\langle \Delta \rangle \geq 0$) as compared to standard Abelian transport { (see Supplemental Material, Sec.~\ref{app:derInterpretation} for more details)}. 

	Finally, since the XFT in Eq.~\eqref{eq:fluctuationsAverage} is a so-called strong fluctuation theorem~\cite{Luposchainsky13}, we can directly apply the results in Ref.~\cite{Hasegawa19,Timpanaro19} to derive an associated Thermodynamic Uncertainty Relation:
	\begin{equation} \label{eq:FTUR}
		\frac{{\rm Var}[\Delta Q_j]}{\langle \Delta Q_j \rangle^2} \geq \frac{2}{e^{\sum_i \delta \lambda_i \langle \Delta Q_i \rangle + \langle \Delta \rangle} - 1},
	\end{equation}
	which is valid for any current of non-commuting charges $\Delta Q_j$ with $j=1, ..., N$ (the proof is given in the Supplemental Material, Sec.~\ref{app:derTUR}). We note that, whenever $\langle \Delta \rangle \geq 0$, the TUR~\eqref{eq:FTUR} predicts the possibility of improved current precision as compared to the one allowed by the TUR for commuting charges ($\Delta = 0$). This reveals that the role of non-commutativity in providing a thermodynamic advantage goes beyond the case of average currents.

	{\it Illustrative Example.} We exemplify our findings with the interaction between { two} spin-1/2 particles with { local Hamiltonians {\color{black} $H_A = \hbar \omega \id^A$ and $H_B = \hbar\omega \id^B$ (for simplicity, we set $\hbar \omega = 1$)}\footnote{{\color{black} It should be noticed that for qubit systems hosting two non-commuting charges, the choice of a fully degenerate system Hamiltonian is necessary, as we need to have $[H_K, Q_1^K] =[H_K, Q_2^K] = 0$  while $[Q_1^K,Q_2^K] \neq 0$ (where $K = A, B$). This is only possible if the system Hamiltonian has at least one eigenvalue that has degeneracy $2$, i.e., for qubits, $H_K \propto \id_K$.}} coming from baths with two non-commuting charges corresponding to {{\color{black} $Q_1^{K}= \sigma_z^K$} and $Q_2^{K}=\sigma_x^K$ for $K = A, B$}, as described by the generalized Gibbs state {:
			\begin{align}
				\pi_{\lambda^{\bf K}}^{K} := \frac{e^{-\beta^K \sigma_z^K -\chi^K \sigma_x^{K}}}{Z_K} \,,
			\end{align}
			with $\sigma_z^K$ and $\sigma_x^K$ being the usual Pauli operators acting on each spin $K=A,B$. The corresponding affinities are $\lambda_1^K = \beta^K$ and $\lambda_2^K = \chi^K$}. 
		In the collisional picture of Fig.~\ref{fig:framework} a particle from each bath (with affinities $\lambda^{\bf A} = \{\beta^A,\chi^A\}$ and $\lambda^{\bf B}= \{\beta^B,\chi^B\}$) is measured in the basis of $\mathcal{H}_i=\beta^i \sigma_z^{(i)} + \chi^i \sigma_x^{(i)}$ with $i=A,B$, interacts through the unitary $U$, and finally returns to its own bath after being measured again. The constraint in Eq.~\eqref{eq:commutation} singles out generalized swaps as the unique allowed interaction Hamiltonian (see Supplemental Material, Sec.~\ref{app:int} for details { about this result, and in particular see Eq. (S44) for the explicit expression). This type of dynamics can be experimentally implemented with trapped ions through laser-induced entangling interactions~\cite{Kranzl23}.}
		
		We are now ready to compute and test the currents probability distribution $P(\Delta\sigma_z,\Delta\sigma_x;\Delta)$ to verify the XFT in Eq.~\eqref{eq:XFT}, and, in particular, it is interesting to compare it with the marginal distribution
		$P(\Delta\sigma_z,\Delta\sigma_x):= \int\de\Delta \;P(\Delta\sigma_z,\Delta\sigma_x;\Delta)\,$.
		Indeed, as exemplified in Fig.~\ref{fig:simulations}.a), while the integral XFT in Eq.~\eqref{eq:Jarzynski} is always verified, the standard fluctuation theorem can be broken, i.e. 
		$\average{e^{-\sum_{i} {\delta\lambda_i \Delta Q_i}}}{} \neq 1. $
		Moreover, since $\Delta_\gamma$ is different from zero for all trajectories if the conserved charges do not commute among each other; then eventual breakdown of the standard fluctuation theorem would witness the effective non-commutation of the charges involved in the process.
		
		The behavior of the second-law inequality Eq.~\eqref{eq:secondlaw} with and without quantum correction is plotted in Fig.~\ref{fig:simulations}.b). There, we can appreciate regions with apparent violations of the second law, whenever the quantum correction term $\langle \Delta \rangle$ is not incorporated. These regions are highlighted with a shaded area and are clearly related with a large instability of the standard integral XFT (see Fig.~\ref{fig:simulations}.a).
		
		It is interesting to point out that { the anomalous regions} 
		can also lead to inversions in the currents with respect to the expected spontaneous flow, a purely quantum effect. To be more precise, we observe an inversion of the standard behavior of both currents $\average{\Delta \sigma_z}{}$ and $\average{\Delta \sigma_x}{}$ against their respective affinity differences. This is opposed to the case where the inversion with respect to the affinity gradient occurs only in one of the two currents, whereas the standard flow in the other current compensates for the first (i.e., when there is a trade-off between two resources~\cite{Guryanova16}). This behavior is possible only due to a non-zero term $\average{\Delta }{}$, and is thus a direct consequence of the non-commutativity of the charges. In Fig.~\ref{fig:simulations}.c) we show two such possibilities, where $\beta^A\equiv\beta^B$ and $\beta^A<\beta^B$. 
		
		To be more precise, {\color{black} the natural direction of the current $\average{\Delta \sigma_z}{}$ is from a smaller affinity to a larger one (in analogy to what would happen for inverse temperatures and energy currents). This means that for $\beta^A> \beta^B$ one would expect $\average{\Delta \sigma_z}{} \geq 0$, and the other way round. Then, an inversion of $\average{\Delta \sigma_z}{}$ occurs if $\average{\Delta \sigma_z}{}\cdot\delta\beta<0$. 
			Similarly, an inversion of the the current $\average{\Delta \sigma_x}{}$ occurs if $\average{\Delta \sigma_x}{}\cdot\delta\chi<0$. In Fig.~\ref{fig:simulations}.c) we observe some parameter regions for which the expected current in $\langle \Delta \sigma_z \rangle$ (that is $\beta^A < \beta ^B$  and $\langle \Delta \sigma_z \rangle \leq 0$, analogous to an energy current from a hot to a cold bath) is used to invert the current $\langle \Delta\sigma_x \rangle$ against the bias ($\chi_A < \chi_B$) (e.g., in the central, unshaded area of the bottom plot in Fig.~\ref{fig:simulations}.c)}. However, more striking situations are highlighted in the orange shaded areas of the plots, where we see that both current inversions can happen at the same time, exemplifying the possibility of using non-commutative charges to generate currents that are opposite to the affinities gradient.
		
		
		
		
		{\it Conclusions.} We have seen that the statistics of charge currents in non-Abelian transport obey universal nonequilibrium relations which can be considered a generalization of detailed and integral XFT to non-commuting charges, that is Eqs.~\eqref{eq:fluctuationsAverage} and \eqref{eq:Jarzynski}, together with inequalities~\eqref{eq:secondlaw} and~\eqref{eq:FTUR}. These provide a refinement of the second law for non-Abelian transport with a quantum correction term that allows classically forbidden behavior such as the inversion of all currents against their affinity biases or enhanced precision in the current fluctuations. 
		
		For the derivation of the XFT we used a collisional transport framework following Ref.~\cite{Manzano22}, over which we introduced a noninvasive two-point measurement scheme which suits typical transport setups where the two baths exchanging charges are effectively uncorrelated to each other. {\color{black} Our results have universal character, being this setup equivalent in generality and scope to the ones typically employed for the derivation of XFTs (see e.g. Ref.~\cite{Campisi11})~\footnote{{\color{black} We also expect that our results could also be derived from an appropriate full counting statistics scheme in the context of continuously monitored systems.}}}. Extensions of our results might be considered by incorporating initial (classical) correlations between the baths similarly to Ref.~\cite{Jevtic15}, or quantum ones following Ref.~\cite{Micadei20}, {which may also lead to anomalous heat flow~\cite{Comar25,Guan25}}. A comparison with alternative methods based on quasi-probability distributions is also desirable~\cite{Levy20}. Experimental tests of the XFT, the TUR and the second-law inequality for non-Abelian transport might be implemented for spin chains on trapped ion platforms~\cite{Kranzl23} or for squeezed thermal reservoirs~\cite{Klaers17}.   
		
		{\it Acknowledgements.} M.S. acknowledges support from the Spanish Agencia Estatal de Investigacion through the grants ``IFT Centro de Excelencia Severo Ochoa CEX2020-001007-S” and ``PCI2024-153448'’, funded by MICIU/AEI/10.13039/501100011033 and co-funded by the EU. This project was funded within the QuantERA II Programme that has received funding from the EU’s H2020 research and innovation programme under the GA No 101017733.
		G.M. acknowledges financial support from the Ramón y Cajal program RYC2021-031121-I funded by MICIU/AEI/10.13039/501100011033 and European Union NextGenerationEU/PRTR, the CoQuSy project PID2022140506NB-C21 and C22, and the María de Maeztu project CEX2021-001164-M for Units of Excellence, funded by MICIU/AEI/10.13039/501100011033/FEDER, UE.

			\bibliography{refs.bib}

		\clearpage
		\onecolumngrid
		
		\vspace{40pt}
		\begin{center}
			\textbf{\large Supplemental Material to ``Exchange Fluctuation Theorem for Non-Commuting Charges"}
		\end{center}

		\setcounter{equation}{0}
		\setcounter{figure}{0}
		\setcounter{table}{0}
		\setcounter{page}{1}
		\makeatletter
		\renewcommand{\thesection}{S\arabic{section}}
		\renewcommand{\theequation}{S\arabic{equation}}
		\renewcommand{\thefigure}{S\arabic{figure}}

		The Supplemental Material contains detailed proofs and further details on the derivation of the main results. In Sec.~\ref{app:derSelectionRule} we provide a proof of the charges conservation relation in  Eq.~\eqref{eq:selectionRuleCompact} of the main text. Sec.~\ref{app:derTimeReversal} is devoted to provide details of the definition of the backward process and the time-reversed trajectories, and provide details on the derivation of Eq.~\eqref{eq:fluctuationTrajectories} of the main text. In Sec.~\ref{app:derFluctuations} we provide the proof of the detailed fluctuation theorem for currents of non-commuting charges presented in Eq.~\eqref{eq:Jarzynski} of the main text. Sec.~\ref{app:derStatistical} is devoted to the derivation of the integral fluctuation theorem and highlights the connexion of our results with statistical quantifiers of interest. Sec.~\ref{app:derTUR} we provide a proof of the Thermodynamic Uncertainty Relation, Eq.~\eqref{eq:FTUR} of the main text. Finally, details on how the separate conservation of (non-commuting) spin-1/2 components singles out the generalized swaps as the unique allowed interaction Hamiltonian are provided in Sec.~\ref{app:int}.
		
		\section{Derivation of the conservation relation in Eq.~(5)}\label{app:derSelectionRule}
		
		The charges conservation relations in Eq.~\eqref{eq:commutation} from the main text imply that the interaction Hamiltonian satisfies $[H_{\rm int}, Q^A_i + Q^B_i]=0$ for all $i= 1, ..., N$. If we multiply each equation by their corresponding affinity $\lambda_i^B$, and we sum over different charges, these then yield:
		\begin{align}
			0 = \sum_i \; \lambda^B_i \, \sqrbra{{H}_{\rm int}, Q_i^A + Q_i^B} = [{H}_{\rm int},\mathcal{H}_A+\mathcal{H}_B] - \sum_i \;\delta\lambda_i \sqrbra{{H}_{\rm int}, Q_i^A}\,,
		\end{align}
		where we used the definition $\mathcal{H}^X= \sum_i \lambda_i^X Q_i^X$ for $X= A, B$. Now, sandwiching the expression above with the states $\ket{h_{m,\mu}}$ and $\ket{h_{n,\nu}}$, we obtain the relation:
		\begin{align}
			&\qquad \bra{h_{m,\mu}} \sqrbra{H_{\rm int},\mathcal{H}_A+ \mathcal{H}_B}\ket{h_{n,\nu}} = \sum_i \;\delta\lambda_i \bra{h_{m,\mu}}\sqrbra{H_{\rm int}, Q_i^A}\ket{h_{n,\nu}},
		\end{align}
		which, thanks to the product structure of the eigenstates of $\mathcal{H}_A + \mathcal{H}_B$, can be further rewritten as:
		\begin{align}
			&\bra{h_{m,\mu}} H_{\rm int}\ket{h_{n,\nu}} \norbraB{(h^A_m-h^A_n)+(h^B_\mu-h^B_\nu)}=  -\sum_i \;\delta\lambda_i \bra{h_{m,\mu}}\sqrbra{H_{\rm int}, Q_i^A}\ket{h_{n,\nu}} \,.\label{eq:selectionRule}
		\end{align}
		Provided the overlap $|\!\bra{h_{m,\mu}} H_{\rm int} \ket{h_{n,\nu}}\!|^2$ is non-zero, Eq.~\eqref{eq:selectionRule} implies the relation: 
		\begin{align}
			\Delta h^A (n\rightarrow m) + \Delta h^B (\nu\rightarrow \mu)  = -\sum_i \;\delta\lambda_i \frac{\bra{h_{m,\mu}}\sqrbra{H_{\rm int}, Q_i^A}\ket{h_{n,\nu}}}{\bra{h_{m,\mu}} H_{\rm int}\ket{h_{n,\nu}}}, \label{eq:protoresult}
		\end{align}
		where we used the notation $\Delta h^A({n\rightarrow m}) :=(h^A_m-h^A_n)$, and similarly for $\Delta h^B (\nu\rightarrow \mu)$.  We can further expand the left hand side of the above equation to obtain:
		\begin{align}
			&-\sum_i \;\delta\lambda_i \frac{\bra{h_{m,\mu}}\sqrbra{H_{\rm int}, Q_i^A}\ket{h_{n,\nu}}}{\bra{h_{m,\mu}} H_{\rm int}\ket{h_{n,\nu}}} = 	\sum_{i,k} \;\delta\lambda_i \norbra{\bra{h_m^A}Q_i^A\ket{h_{k}^A}\frac{\bra{h_{k,\mu}}H_{\rm int} \ket{h_{n,\nu}}}{\bra{h_{m,\mu}} H_{\rm int}\ket{h_{n,\nu}}}-\frac{\bra{h_{m,\mu}}H_{\rm int} \ket{h_{k,\nu}}}{\bra{h_{m,\mu}} H_{\rm int}\ket{h_{n,\nu}}}\bra{h_k^A} Q_i^A\ket{h_{n}^A}} \nonumber \\
			&= \sum_{i} \;\delta\lambda_i\, \Delta Q_i({n\rightarrow m}) + \Delta(\gamma) 
		\end{align}    
		where we have divided the diagonal and off-diagonal contributions as $\Delta Q_i ({n\rightarrow m}) := \bra{h_m^A}Q_i^A\ket{h_{m}^A}-\bra{h_n^A}Q_i^A\ket{h_{n}^A}$ and the stochastic correction term:
		\begin{align}
			\Delta(\gamma) := \sum_{i,k\neq m} \;\delta\lambda_i \norbra{\bra{h_m}Q_i^A\ket{h_{k}}\frac{\bra{h_{k,\mu}}H_{\rm int} \ket{h_{n,\nu}}}{\bra{h_{m,\mu}} H_{\rm int}\ket{h_{n,\nu}}}} - \sum_{i,k\neq n}\delta\lambda_i\norbra{\bra{h_k}Q_i^A\ket{h_{n}}\frac{\bra{h_{m,\mu}}H_{\rm int} \ket{h_{k,\nu}}}{\bra{h_{m,\mu}} H_{\rm int}\ket{h_{n,\nu}}}}\,,\label{eq:S6}
		\end{align}
		which was introduced in Eq.~\eqref{eq:DeltaExpr} in the main text. Using these identifications we can now rearrange Eq.~\eqref{eq:protoresult} to obtain:
		\begin{equation} \label{eq:conservation}
			\Delta h^A({n\rightarrow m}) + \Delta h^B({\nu\rightarrow \mu}) = \sum_{i} \;\delta\lambda_i\, \Delta Q_i({n\rightarrow m}) + \Delta({\gamma})\,
		\end{equation}
		proving Eq.~\eqref{eq:selectionRuleCompact} of the main text. This equation can be interpreted as follows: each $\Delta Q_i ({n\rightarrow m})$ corresponds to the stochastic exchange in the charge $Q_i$ between the beginning and the end of the collision (as measured in system $A$). Moreover, $\Delta({\gamma})$ is a correction solely arising from the non-commutativity of the conserved charges. 
		
		This can be understood as follows: suppose that the charges would commute, that is for all $i$ and $j$, $[Q_i, Q_j]= 0$. Then this would also imply that $[\mathcal{H},Q_i]=0$. Since this is equivalent to the existence of a shared eigenbasis,  we also have that all off-diagonal terms of the form $\bra{h_l^A} Q_i^A \ket{h_k^A}$ are zero in this case, making Eq.~\eqref{eq:S6} identically zero.
		
		It should also be noticed that $\Delta({\gamma})$ is real, since $\Delta Q_i({n\rightarrow m})$ and $\Delta h^A({n\rightarrow m}) + \Delta h^B({\nu\rightarrow \mu})$ are all manifestly real quantities. This is a general fact about commutators. Indeed, let $A$ and $B$ be two self-adjoint operators, and let $[A,B] = [\bar{A}, B]$, where $\bar{A}$ is self-adjoint as well. Then, when expanding in the eigenbasis of $A:= \sum_{A_i} \;A_i \ket{A_i}$, we can repeat the same steps above to obtain:
		\begin{align}
			A_i-A_j =\frac{\sandwich{A_i}{[\bar{A},B]}{A_j}}{\sandwich{A_i}{B}{A_j}}\,.
		\end{align}
		Since the quantity on the left is real, the same has to hold for the quantity on the right as well. This proves that the right hand side of Eq.~\eqref{eq:protoresult} is real.
		
		Finally, consider the correction associated to the backward trajectory $\tilde\gamma = \{(m ,\mu), (n, \nu) \}$ associated to the inverse transition $\ket{h_{m, \mu}} \rightarrow \ket{h_{n, \nu}}$. This satisfies $\Delta({\tilde\gamma}) = -\Delta({\gamma})$. Indeed, first notice that $\Delta h^A({n\rightarrow m})=-\Delta h^A({m\rightarrow n})$ (and similarly for $\Delta h^B({\nu\rightarrow \mu})$). Then, the correction term reads:
		\begin{align}
			& \Delta({\tilde\gamma}) =\sum_{i,k\neq n} \;\delta\lambda_i \norbra{\bra{h_n^A}Q_i^A\ket{h_{k}^A}\frac{\bra{h_{k,\nu}}H_{\rm int} \ket{h_{m,\mu}}}{\bra{h_{n,\nu}} H_{\rm int}\ket{h_{m,\mu}}}} \;- \sum_{i,k\neq m}\delta\lambda_i\norbra{\bra{h_k^A}Q_i^A\ket{h_{m}^A}\frac{\bra{h_{n,\nu}}H_{\rm int} \ket{h_{k,\mu}}}{\bra{h_{n,\nu}} H_{\rm int}\ket{h_{m,\mu}}}} =\nonumber \\
			&=\sum_{i,k\neq n} \;\delta\lambda_i \norbra{\bra{h_k^A}Q_i^A\ket{h_{n}^A}\frac{\bra{h_{m,\mu}}H_{\rm int}\ket{h_{k,\nu}}}{\bra{h_{m,\mu}} H_{\rm int}\ket{h_{n,\nu}}}} \;- \sum_{i,k\neq m}\delta\lambda_i\norbra{\bra{h_m^A}Q_i^A\ket{h_{k}^A}\frac{\bra{h_{k,\mu}}H_{\rm int}\ket{h_{n,\nu}}}{\bra{h_{m,\mu}} H_{\rm int}\ket{h_{n,\nu}}}}  = -\Delta(\gamma)\,,
		\end{align}
		where in the second line we took the complex conjugate of the first expression, and used the fact that $\Delta({\tilde\gamma})$ is real to equate the two expression. This proves the claim.
		
		\section{Time-reversed trajectories and backward process}\label{app:derTimeReversal}
		Here we provide a retailed definition of the backward process standardly used in the context of quantum and stochastic thermodynamics~\cite{Manzano18b,Manzano22b,Campisi11,Esposito14}, which consist on the implementation of the time-reversed protocol or sequence of steps defined in the collisional model introduced in the main text. Moreover, for the sake of generality we may consider charges that may be either even or odd under time-reversal, that is $\Theta Q_i \Theta^{-1} = \pm Q_i$, for which we need to introduce the anti-unitary time-reversal operator in quantum mechanics $\Theta$, and consider all measurements in the two-point-measurement scheme  associated to the transformed (time-reversal) observables $\Theta \mathcal{H}_X \Theta^{-1} = \sum_i \lambda_i^X \Theta Q_i^X \Theta^{-1}$.  
		
		Using the above prescriptions, the backward process of the collisional model reads:
		\begin{enumerate}
			\item An environmental unit (molecule) is taken from each bath, leading to an initial product state $\tilde \rho_0 = \Theta \left(\pi^A_{\lambda^A} \otimes \pi^B_{\lambda^B} \right) \Theta^{-1}$, and is measured by projectors $\Theta \Pi^A \Theta^{-1}$ and $\Theta \Pi^B \Theta^{-1}$ in the eigenbasis of $\Theta \mathcal{H}_{A} \Theta^{-1}$ and $\Theta \mathcal{H}_{B} \Theta^{-1}$ respectively;
			\item The two units interact through the (backward) unitary operation $\tilde{U} = \mathbb{T} e^{-i \int_0^\tau dt \Theta H_{\rm int}(\tau - t)\Theta^{-1}}$, where $H_{\rm int}$ is the same interaction Hamiltonian as in the original (forward) process that couples the two baths during the interaction time $\tau$;
			\item The units are measured again in the eigenbasis of $\Theta \mathcal{H}_{A} \Theta^{-1}$ and $\Theta \mathcal{H}_{B} \Theta^{-1}$ respectively, after which they return to their respective baths.
		\end{enumerate}
		
		It is important to notice that in the backward process the equivalent charge-preserving condition for the interaction between the bath units for all charges $i$ is verified, as we have that:
		\begin{align}
			\sqrbra{\tilde{U}, \Theta \left( Q^A_i + Q^B_i \right)\Theta^{-1}} =\Theta\sqrbra{{U}, \left( Q^A_i + Q^B_i \right)}\Theta^{-1} = 0\,,\label{eq:commutationback}
		\end{align}
		where the last equality follows directly from Eq.~\eqref{eq:commutation} of the main text. Given that the switching protocol of the interaction Hamiltonian $H_{\rm int}$ is symmetric under time reversal, the condition above can also be expressed as $[\Theta H_{\rm int} \Theta^{-1}, \Theta \left(Q^A_i + Q^B_i \right) \Theta^{-1}]=0$, in terms of the interaction Hamiltonian of the collisions $H_{\rm int}(t)$.
		
		Let's now calculate the probability in the backward process of a trajectory $\tilde \gamma = \{(m, \mu), (n,\nu) \}$, that is, the inverse trajectory of the trajectory $\gamma = \{(n, \nu), (m,\mu) \}$ discussed in the main text. In such a trajectory, the results of the first measurement in the backward protocol described above are $h_m^A$ and $h_\mu^B$, while after interaction of the environmental units through unitary $\tilde{U}$, the second measurement gives results $h_n^A$ and $h_\nu^B$. The probability of the inverse trajectory $\tilde \gamma$ in the backward process reads:
		\begin{equation}
			\tilde{\mathbb{P}}(\tilde{\gamma}) = \frac{e^{-(h_m^A + h_\mu^B)}}{Z} | \langle h_{n, \nu}|\Theta^{-1} \tilde U \Theta | h_{m,\mu} \rangle|^2. \label{eq:backprob}
		\end{equation}
		Notice that above we denoted the probability in the backward process $\tilde{\mathbb{P}}$ (with tilde) to distinguish it from the one calculated in the forward process $\mathbb{P}$ (without tilde). However, it follows that the probability $\tilde{\mathbb{P}}$ equals $\mathbb{P}$ in the important case in which $\Theta H_{\rm int} \Theta^{-1} = H_{\rm int}$ and the switching protocol is time-symmetric (we also remark that $H(0)= H(\tau) = 0$), which implies $\tilde{U} = U$, and the charges are even under time-reversal ($\Theta Q_i^X \Theta^{-1} = Q_i^X $), which implies that the measurement projectors in backward process $\Theta \Pi^X \Theta^{-1}$ are equal to the ones appearing in the original (forward process). In that case:
		\begin{equation}
			| \langle h_{n, \nu}|\Theta^{-1} \tilde U \Theta | h_{m,\mu} \rangle = | \langle h_{n, \nu}| \tilde U | h_{m,\mu} \rangle = | \langle h_{n, \nu}| U | h_{m,\mu} \rangle = | \langle h_{n, \nu}| U^\dagger | h_{m,\mu} \rangle,       
		\end{equation}
		and the above expression in Eq.~\eqref{eq:backprob} coincides with the probability of the inverse trajectory $\tilde \gamma$ in the forward process, that is, $ \tilde{\mathbb{P}}(\tilde{\gamma}) = \mathbb{P}(\gamma)$.
		
		The relation in Eq.~\eqref{eq:trajprobs} of the main text then follows by simply comparing the probabilities of trajectory $\gamma$ in the forward process and trajectory $\tilde{\gamma}$ in the backward process. In the general case it reads:
		\begin{align} \label{eq:trajprobsapp}
			\frac{\mathbb{P}(\gamma)}{\tilde{\mathbb{P}}(\tilde{\gamma})} &= \frac{e^{-(h_n^A+h_\nu^B)}}{e^{-(h_m^A+h_\mu^B)}}\;\frac{|\!\sandwich{h_{m,\mu}}{U}{h_{n,\nu}}|^2}{|\!\sandwich{h_{n,\nu}}{\Theta^{-1} \tilde{U} \Theta}{h_{m,\mu}}|^2} = \frac{e^{-(h_n^A+h_\nu^B)}}{e^{-(h_m^A+h_\mu^B)}}\;\frac{|\!\sandwich{h_{m,\mu}}{U}{h_{n,\nu}}|^2}{|\!\sandwich{h_{n,\nu}}{U^\dagger}{h_{m,\mu}}|^2} = e^{\Delta h^A({n\rightarrow m}) + \Delta h^B({\nu\rightarrow \mu}) } \,,
		\end{align}
		where in the second equality we used the micro-reversibility principle for unitary evolutions $\Theta^{-1} \tilde{U} \Theta = U^\dagger$, and the last equality follows from the fact that the conditional probabilities in the numerator and denominator are equal. The relation reported in Eq.~\eqref{eq:trajprobs} of the main text correspond to the particular case for which $\tilde{\mathbb{P}}= \mathbb{P}$.
		
		\section{Derivation of the currents fluctuation theorem}\label{app:derFluctuations}
		
		{ We begin by proving that $P(\Delta Q_1,\dots, \Delta Q_N; \Delta)$ is a bona-fide probability distribution. This trivially follows from its definition and the fact that $\mathbb{P}(\gamma) $ is a bona-fide probability as well. The latter fact can be verified since by definition for all trajectories $\mathbb{P}(\gamma)\geq 0$. Moreover, we have the normalization:
			\begin{align}
				\sum_{\gamma} \;\mathbb{P}(\gamma) &= \sum_{m,\mu,n,\nu}\;\frac{e^{-(h_n^A+h_\nu^B)}}{Z}\;|\!\sandwich{h_{m,\mu}}{U}{h_{n,\nu}}|^2 = 
				\\
				&=\sum_{n,\nu}\;\frac{e^{-(h_n^A+h_\nu^B)}}{Z}\;\Tr{U^\dagger\norbra{\sum_{m,\mu}\ket{h_{m,\mu}}\bra{h_{m,\mu}}}U\ket{h_{n,\nu}}\bra{h_{n,\nu}}} =
				\\
				&=\sum_{n,\nu}\;\frac{e^{-(h_n^A+h_\nu^B)}}{Z} = \frac{Z}{Z} =1\,,
			\end{align}
			where in the second line we have used the decomposition of the identity in the eigenbasis $\ket{h_{m,\mu}}$, and finally the definition of $Z$. Then, it follows from the definition of $P(\Delta Q_1,\dots, \Delta Q_N; \Delta)$:
			\begin{align}
				&P(\Delta Q_1,\dots, \Delta Q_N; \Delta) =\sum_{\gamma} \; \mathbb{P}(\gamma) \, \delta\norbra{\Delta Q_1- \Delta Q_1({n\rightarrow m})} \dots \delta\norbra{\Delta Q_N- \Delta  Q_N({n\rightarrow m})}~\delta\norbra{\Delta - \Delta(\gamma)}\,,
			\end{align}
			that it is automatically positive, since it is the sum of positive terms. Moreover, carrying out the sum over $\Delta Q_1\,,\dots\Delta Q_N$ and $\Delta$, takes care of the delta-functions. This shows that it is also normalized, so it is a bona-fide probability distribution.
		}
		
		First, notice that the probability distribution $P_B(\Delta Q_1,\dots, \Delta Q_N; \Delta)$ associated with the backward trajectories $\tilde{\gamma}$ satisfies the equation:
		\begin{align}
			&P_B(-\Delta Q_1,\dots, -\Delta Q_N; -\Delta) = \sum_{\gamma} \; \tilde{\mathbb{P}}(\tilde{\gamma}) \, \delta\norbra{\Delta Q_1+ \Delta Q_1({m\rightarrow n})}\dots \delta\norbra{\Delta Q_N+ \Delta Q_N(m\rightarrow n)} \delta\norbra{\Delta + \Delta(\tilde{\gamma})}= \nonumber \\
			&=\sum_{\gamma} \; \mathbb{P}(\gamma) e^{-\sum_{i} \delta\lambda_i \Delta Q_i({n\rightarrow m}) - \Delta({\gamma})} \, \delta\norbra{\Delta Q_1- \Delta Q_1({n\rightarrow m})}\dots \delta\norbra{\Delta Q_N- \Delta Q_N({n\rightarrow m})}\delta\norbra{\Delta - \Delta_\gamma} = \\
			&=P(\Delta Q_1,\dots, \Delta Q_N; \Delta) \, e^{-\sum_{i} \delta\lambda_i \Delta Q_i\,-\,\Delta}\,, \nonumber
		\end{align}
		where in the second line we used Eq.~\eqref{eq:fluctuationTrajectories}, together with the relations $\Delta Q_i({m\rightarrow n}) = -\Delta Q_i({n\rightarrow m})$ and $\Delta({\tilde\gamma}) = -\Delta(\gamma)$ to change the variables from $\tilde\gamma$ to $\gamma$. Then, reordering we finally obtain:
		\begin{align} \label{eq:detailed}
			\frac{P(\Delta Q_1,\dots, \Delta Q_N; \Delta)}{P_B(-\Delta Q_1,\dots, -\Delta Q_N; - \Delta)} =  e^{\sum_{i} \norbra{\delta\lambda_i \Delta Q_i}\,+\,\Delta}\,,
		\end{align}
		which is a generalized version of Eq.~\eqref{eq:fluctuationsAverage} in the main text to the case in which $\tilde {\mathbb{P}}(\tilde \gamma) \neq \mathbb{P}(\tilde \gamma)$, that is, in the presence of charges or interaction Hamiltonian that are odd under time-reversal. The relation above simplifies if we assume $\tilde {\mathbb{P}}(\tilde \gamma) = \mathbb{P}(\tilde \gamma)$, since in this case we have:
		\begin{align}
			&P_B(-\Delta Q_1,\dots, -\Delta Q_N; -\Delta) = \sum_{\gamma} \; \tilde{\mathbb{P}}(\tilde{\gamma}) \, \delta\norbra{\Delta Q_1+ \Delta Q_1({m\rightarrow n})}\dots \delta\norbra{\Delta Q_N+ \Delta Q_N(m\rightarrow n)} \delta\norbra{\Delta + \Delta(\tilde{\gamma})}= \nonumber \\
			&=\sum_{\gamma} \; \mathbb{P}(\tilde{\gamma}) \, \delta\norbra{\Delta Q_1+ \Delta Q_1({m\rightarrow n})}\dots \delta\norbra{\Delta Q_N+ \Delta Q_N(m\rightarrow n)} \delta\norbra{\Delta + \Delta(\tilde{\gamma})} = P(-\Delta Q_1,\dots, -\Delta Q_N; -\Delta)\,.
		\end{align}
		Then, substituting this relation into Eq.~\eqref{eq:detailed}, we finally obtain Eq.~\eqref{eq:fluctuationsAverage} from the main text.
		
		\section{Integral fluctuation theorem and second-law inequalities} \label{app:derStatistical}
		
		In the following we consider the family of statistical divergences (also called contrast functions) dependent on the function $g(x)$:
		\begin{align} \label{eq:contrast}
			S_g(\rho||\sigma) := \Tr{g(\LL_\sigma\RR_{\rho}^{-1})[\rho]} = \sum_{i,j}\; \rho_i \,g\norbra{\frac{\sigma_j}{\rho_i}}\, |\braket{\sigma_j}{\rho_i}|^2\,,
		\end{align}
		where we used the eigendecomposition $\rho:= \sum_i \rho_i\ketbra{\rho_i}{\rho_i}$ and $\sigma:= \sum_i \sigma_i\ketbra{\sigma_i}{\sigma_i}$, { and we introduced the left and right multiplication operators, $\LL_{\sigma} (X) := \sigma X$ and $\RR_\rho(X):= X\rho$}. In order to verify the standard properties of statistical divergences the function $g(x)$ needs to be matrix convex, satisfy that $g(1) = 0$ is its unique zero, and that, for some arbitrary constant $a$, it holds that $g(x) + a(x-1) \geq 0$~\cite{Scandi25}. We note that for $g(x) = -\log(x)$, we recover the relative entropy, $S(\rho||\sigma) =\Tr{\rho\,(\log\rho- \log \sigma)}$, while other choices may be done that recover the relative entropy variance or the family of $\alpha$-divergences. 
		
		If we now substitute in Eq.~\eqref{eq:contrast} $\sigma:=\rho_0= \pi^A_{\lambda^A} \otimes\pi^B_{\lambda^B}$ and $\rho := U\rho_0 U^\dagger$ it is easy to see that:
		\begin{align}
			S_g(U\pi U^\dagger||\pi) &= \sum_{n,m,\nu,\mu}\; \frac{e^{-(h_n^A+h_\nu^B)}}{Z}\,g\norbra{\frac{e^{-(h_m^A+h_\mu^B)}}{e^{-(h_n^A+h_\nu^B)}}}\;|\!\sandwich{h_{m,\mu}}{U}{h_{n,\nu}}|^2=\nonumber\\
			&= \sum_\gamma \;g\norbraB{e^{-\Delta h^A({n\rightarrow m}) - \Delta h^B({\nu\rightarrow \mu})}}\,\mathbb{P}(\gamma) = \sum_\gamma \;g\norbraB{e^{-\sum_{i} \delta\lambda_i \Delta Q_i({n\rightarrow m})\, -\, \Delta({\gamma})}}\,\mathbb{P}(\gamma)=\label{eq:contrastFunctionRelation}\\
			&=\int \de \Delta Q_1\dots \de \Delta Q_N\de\Delta\; g\norbraB{e^{-\sum_{i} \delta\lambda_i \Delta Q_i\, -\, \Delta}} \, P(\Delta Q_1,\dots, \Delta Q_N; \Delta)\,,\nonumber
		\end{align}
		
		
		
		The relation in Eq.~\eqref{eq:contrastFunctionRelation} allows us to derive an integral fluctuation theorem. Indeed, setting $g(x) =x$, the corresponding contrast function reduces to $S_{x}(\rho||\sigma) = \Tr{\sigma} = 1$. Thus:
		\begin{align}
			\average{e^{-\sum_{i} \norbra{\delta\lambda_i \Delta Q_i}\, -\, \Delta}}{} := \int \de \Delta Q_1\dots \de \Delta Q_N\de\Delta\; e^{-\sum_{i} \delta\lambda_i \Delta Q_i\, -\, \Delta}\, P(\Delta Q_1,\dots, \Delta Q_N; \Delta) = 1\,.
		\end{align}
		This proves the Jarzynski-like version of the XFT in Eq.~\eqref{eq:Jarzynski}.
		Moreover, a well-known consequence of the (standard) integral fluctuation theorems is that statistical violations of the second law are exponentially damped~\cite{Jarzynski11}. Then, it follows from Eq.~\eqref{eq:Jarzynski} that:
		\begin{align}
			P\norbra{\sum_{i} \delta\lambda_i \Delta Q_i +\Delta <-\zeta} \leq \;e^{-\zeta}\,,\label{eq:s20}
		\end{align}
		where we implicitly defined the quantity on the left-hand side to be the probability, according to $P$, to observe a total value of $\sum_{i} {\delta\lambda_i \Delta Q_i}+\Delta$ less than $-\zeta$ for any real scalar $\zeta$. Here we see that the presence of the quantum correction allows for the probability distribution of $\sum_{i} {\delta\lambda_i \Delta Q_i}$ to be more skewed towards negative values as compared to the case for commuting charges. Both relations show that the presence of the quantum correction term $\Delta$ allows for bigger ``violations" as measured by the sole variation of the charges.
		
		For completeness, we include here a proof of Eq.~\eqref{eq:s20}. This can be verified by the chain of inequalities:
		\begin{align}
			&P\norbra{\sum_{i} \delta\lambda_i \Delta Q_i +\Delta <-\zeta} = \int_{-\infty}^{-\zeta} \de x\; \delta(x - \sum_{i} \delta\lambda_i \Delta Q_i -\Delta)\int \de \Delta Q_1\dots \de \Delta Q_N\de\Delta\; P(\Delta Q_1,\dots, \Delta Q_N; \Delta)=\nonumber \\
			&\qquad=\int_{-\infty}^{-\zeta} \de x\; \delta(x - \sum_{i} \delta\lambda_i \Delta Q_i -\Delta)\int \de \Delta Q_1\dots \de \Delta Q_N\de\Delta\; P_B(-\Delta Q_1,\dots, -\Delta Q_N; -\Delta)e^{\sum_{i} \delta\lambda_i \Delta Q_i\,+\,\Delta}\leq\\
			&\qquad\qquad\leq e^{-\zeta}\;\int \de \Delta Q_1\dots \de \Delta Q_N\de\Delta\; P_B(-\Delta Q_1,\dots, -\Delta Q_N; -\Delta)  \leq \;e^{-\zeta}\nonumber\,.
		\end{align}
		
		The second-law inequality in Eq.~\eqref{eq:secondlaw} of the main text can be derived either by applying Jensen's inequality to the integral fluctuation theorem, of by choosing $g(x) = -\log(x)$ in Eq.~\eqref{eq:contrastFunctionRelation}, which leads to $S(U\rho_0 U^\dagger||\rho_0) = \average{\sum_{i} \delta\lambda_i \Delta Q_i +\, \Delta}{}\ \geq 0 $. In any case, we recover Eq.~\eqref{eq:secondlaw} in the main text, namely:
		\begin{align}
			\sum_{i} \delta\lambda_i \langle \Delta Q_i \rangle \geq - \langle \Delta \rangle
		\end{align}
		where the average is taken with respect to $P(\Delta Q_1,\dots, \Delta Q_N; \Delta)$. 
		

		{
			\section{Interpretation of non-Abelian quantum correction from entropy production} \label{app:derInterpretation}
			
			The quantity $\Delta$ in Eq.~\eqref{eq:DeltaExpr} is central to this work, since it separates non-Abelian transport from the Abelian case. Indeed, if the charges commute between them, we have $\Delta(\gamma) = 0$ for any trajectory $\gamma = \{ (n, \nu), (m, \mu) \}$ (see Sec.~\ref{app:derSelectionRule}). Perhaps the most intuitive way to understand this quantity is as the difference:
			\begin{equation}
				\Delta = \Delta h^A + \Delta h^B - \sum_i \delta \lambda _i Q_i, 
			\end{equation}
			c.f. Eq.~\eqref{eq:selectionRuleCompact} in the main text. Here, the first term $\Delta h_1 + \Delta h_2$ can be interpreted as the information-theoretical definition of entropy production (EP) in the process (from Eq.~\eqref{eq:trajprobs} of the main text):
			\begin{equation}
				S_{\textrm{tot}}(\gamma) = \ln \left(\frac{\mathbb{P}(\gamma)}{\mathbb{P}(\tilde{\gamma})}\right) = \Delta h(n\rightarrow m)^A + \Delta h(\nu \rightarrow \mu)^B,
			\end{equation}
			where we recall that $\Delta h^A$ and $\Delta h^B$ are the stochastic changes in the (inclusive) weighted sum of the charge operators, $\mathcal{H}^X = \sum_i \lambda_i Q_i^X$ for $X=A,B$. The above entropy production fully captures the irreversibility of the process (likelihood of forward and time-reversed trajectories) and is in line with the derivations of Ref~\cite{Manzano18b}. 
			
			At the average level, the correction term reads:
			\begin{equation} \label{eq:diff}
				\langle \Delta \rangle = \langle S_{\textrm{tot}} \rangle - \sum_i \delta \lambda_i \langle \Delta Q_i \rangle,
			\end{equation}
			which can now be interpreted as the difference between two notions of entropy production.
			The first term is the average informational entropy production introduced above, which becomes a Kullback-Leibler divergence:
			\begin{equation} \label{eq:Stot}
				\langle S_{\textrm{tot}} \rangle = D_{\textrm{KL}}[\mathbb{P}(\gamma) || \mathbb{P}(\tilde{\gamma})] = \langle \Delta h^A \rangle + \langle \Delta h^B\rangle \geq 0,
			\end{equation}
			with $\langle \Delta h^A \rangle = \Tr{(U\rho_0 U^\dagger - \rho)} \mathcal{H}_A]$ the average change in the operator $\mathcal{H}_A$, and analogously for $\langle \Delta h^B \rangle$. The inequality above follows from the non-negativity of the Kullback-Leibler divergence, $D_{\textrm{KL}}(P || Q) \geq 0$ for any two probability distributions $P$ and $Q$~\cite{Cover99}. Interestingly, we see from Eq.~\eqref{eq:Stot} that the irreversibility of the process is related to the charge currents when no measurements are made: 
			\begin{equation} \label{eq:currentsnomeas}
				\langle S_{\textrm{tot}} \rangle = \langle \Delta h^A \rangle + \langle \Delta h^B \rangle = \sum_{i} \delta \lambda_i \Tr{Q_i^A (U \rho_0 U^\dagger - \rho_0)}.
			\end{equation}
			
			On the other hand, the second term in Eq.~\eqref{eq:diff}, $\sum_i \delta \lambda_i \langle \Delta Q_i \rangle$, is the standard expression for the (thermodynamic) entropy production~\cite{Landi21}, constructed from the measured currents, that is, $\langle \Delta Q_i \rangle = \Tr{Q_i^A (\rho_1  - \rho_0)}$, with $\rho_1 = \sum_{m, \mu} (\Pi_m^A \otimes \Pi_\mu^B) U \rho_0 U^\dagger (\Pi_m^A \otimes \Pi_\mu^B)$ the state of the system after the second measurement. This expression is exactly equal to Eq.~\eqref{eq:currentsnomeas} if the charges commute, i.e. when $[Q_i^A, \mathcal{H}_A] = 0$ for all $i$, but differs otherwise, as $Q_i^A$ and $\mathcal{H}_A$ have in general different basis, and hence the final measurement may affect the statistics of the charges. 
			
			Therefore, the non-Abelian quantum correction essentially captures the non-commutativity of the charges from the effect of the second measurement on the entropy production. In this way, $\langle \Delta \rangle < 0$  when the measured currents lead to an apparent entropy production that is greater than the actual information-theoretical EP, i.e. when the currents suggest a more irreversible behavior than expected. On the other hand, $\langle \Delta \rangle > 0$ becomes positive when the currents lead to a behavior corresponding to a lower EP, hence leading to a paradoxical behavior which seems more reversible than expected, like the inversion of all currents against their bias.
		}
		
		\section{Fluctuation Theorem Uncertainty Relation for Non-Abelian Transport} \label{app:derTUR}
		
		In this last section we provide a detailed proof of the TUR in Eq.~\eqref{eq:FTUR} of the main text. The proof is based on the derivation presented in Ref.~\cite{Hasegawa19} for classical stochastic systems. A key element in the derivation is the detailed fluctuation theorem for the charge currents in its strong version presented in Eqs.~\eqref{eq:fluctuationTrajectories} and Eq.~\eqref{eq:fluctuationsAverage} of the main text, and the fact that the stochastic charge exchanges $\Delta Q_i (n \rightarrow m)$ and the term $\Delta(\gamma)$ are anti-symmetric under time reversal, for trajectories $\tilde{\gamma}=\{(m,\mu),(n, \nu) \}$, which implies that the quantity $\sigma(\gamma) := \sum_i \delta \lambda_i \Delta Q_i(n \rightarrow m) + \Delta(\gamma)$, is also anti-symmetric under time reversal. This means that:
		\begin{equation}
			\Delta Q_i (m \rightarrow n) = - \Delta Q_i (n \rightarrow m) \qquad;\qquad \Delta(\tilde{\gamma})= - \Delta(\gamma) \qquad;\qquad   \sigma(\tilde \gamma) = - \sigma(\gamma)\,.
		\end{equation}
		
		We then consider the probability to observe a change in $\sigma$ and $\Delta Q_j$ for some $j$ in the forward and backward processes. Similarly to Eq.~\eqref{eq:fluctuationsAverage} in the main text, we have the detailed fluctuation relation:
		\begin{align} \label{eq:newFT}
			P(\sigma, \Delta Q_j) &= \sum_{\gamma} \mathbb{P}(\gamma) \delta(\Delta \sigma - \sigma(\gamma) \delta(\Delta Q_j - \Delta Q_j(n \rightarrow m)) = \sum_{\gamma} \mathbb{P}(\tilde{\gamma}) e^{\sigma(\gamma)} \delta(\Delta \sigma - \sigma(\gamma) \delta(\Delta Q_j - \Delta Q_j(n \rightarrow m)) =\nonumber \\ &= e^{\sigma} \sum_{\gamma} \mathbb{P}(\tilde{\gamma}) \delta(\Delta \sigma + \sigma(\tilde{\gamma}) \delta(\Delta Q_j + \Delta Q_j(m \rightarrow n)) = e^{\sigma} P(-\sigma, - \Delta Q_j),
		\end{align}
		which again corresponds to a strong fluctuation theorem. The TUR can now be derived from Eq.~\eqref{eq:newFT} following the same steps as in Ref.~\cite{Hasegawa19} (see also Ref.~\cite{Merhav10}). There, the authors consider the probability density function for positive $\sigma$:
		\begin{equation}
			Q(\sigma, \Delta Q_j) := (1 + e^{-\sigma}) P(\sigma, \Delta Q_j)\,.
		\end{equation}
		It should be noticed that, thanks to Eq.~\eqref{eq:newFT}, $Q(\sigma, \Delta Q_j)$ is normalized. Indeed, we have that:
		\begin{align}
			\int_0^\infty d\sigma \int_{-\infty}^\infty d\Delta Q_j \, Q(\sigma, \Delta Q_j) &= \int_0^\infty d\sigma \int_{-\infty}^\infty d\Delta Q_j (1 + e^{-\sigma}) P(\sigma, \Delta Q_j)= \nonumber \\ &= \int_0^\infty d\sigma \int_{-\infty}^\infty d\Delta Q_j P(\sigma, \Delta Q_j) + \int_0^\infty d\sigma \int_{-\infty}^\infty d\Delta Q_j P(-\sigma, -\Delta Q_j)=  \\ &= \int_0^\infty d\sigma \int_{-\infty}^\infty d\Delta Q_j P(\sigma, \Delta Q_j) + \int_{-\infty}^0 d\sigma \int_{-\infty}^\infty d\Delta Q_j P(\sigma, \Delta Q_j)= 1\,.\nonumber
		\end{align}
		Let us also by $\langle A \rangle_Q$ the average over the distribution $Q$ of a generic functional $A(\sigma, \Delta Q_j)$, i.e.:
		\begin{equation}
			\langle A \rangle_Q = \int_0^\infty d\sigma \int_{-\infty}^\infty d\Delta Q_j \, A(\sigma,\Delta Q_j) \, Q(\sigma, \Delta Q_j)\,.
		\end{equation}
		Interestingly, the average changes in $\langle \sigma \rangle$ and $\langle \Delta Q_j \rangle$ can be expressed as expectations over the distribution $Q$, that is:
		\begin{align}
			\langle \Delta \sigma \rangle &= \int_{-\infty}^\infty d\sigma \int_{-\infty}^\infty d\Delta Q_j \, P(\sigma, \Delta Q_j) \sigma = \int_0^\infty d\sigma \int_{-\infty}^\infty d\Delta Q_j \, P(\sigma, \Delta Q_j) \,\sigma \, (1 - e^{- \sigma})  =\nonumber \\ 
			&= \int_0^\infty d\sigma \int_{-\infty}^\infty d\Delta Q_j \, Q(\sigma, \Delta Q_j) \,\sigma \,  \tanh\left(\frac{\sigma}{2}\right) = \average{ \sigma \, \tanh\left(\frac{\sigma}{2}\right) }{Q}\,;  \label{eq:current1}\\
			\langle \Delta Q_j \rangle &= \int_{-\infty}^\infty d\sigma \int_{-\infty}^\infty d\Delta Q_j \, P(\sigma, \Delta Q_j) \, \Delta Q_j = \int_0^\infty d\sigma \int_{-\infty}^\infty d\Delta Q_j \, P(\sigma, \Delta Q_j) \, \Delta Q_j (1 - e^{- \sigma}) =\nonumber \\ 
			&= \int_0^\infty d\sigma \int_{-\infty}^\infty d\Delta Q_j \, Q(\sigma, \Delta Q_j) \, \Delta Q_j \tanh\left(\frac{\sigma}{2}\right)  =  \average{\Delta Q_j \tanh\left(\frac{\sigma}{2}\right)}{Q}\,.
		\end{align}
		On the other hand, for the squared current $\langle \Delta Q_j^2 \rangle $  we have:
		\begin{align} \label{eq:current2}
			\langle \Delta Q_j^2 \rangle &= \int_{-\infty}^\infty d\sigma \int_{-\infty}^\infty d\Delta Q_j \, P(\sigma, \Delta Q_j) \, \Delta Q_j^2 = \int_0^\infty d\sigma \int_{-\infty}^\infty d\Delta Q_j \, P(\sigma, \Delta Q_j) \, \Delta Q_j^2 (1 + e^{- \sigma}) =\nonumber \\ &=  \int_0^\infty d\sigma \int_{-\infty}^\infty d\Delta Q_j \, Q(\sigma, \Delta Q_j) \, \Delta Q_j^2 = \langle \Delta Q_j^2 \rangle_Q,   
		\end{align}
		that is, averages over $P$ and $Q$ are equal. Then, we can apply Cauchy–Schwarz inequality to get:
		\begin{equation} \label{eq:CS}
			\langle \Delta Q_j\rangle^2 =  \average{\Delta Q_j \tanh\left(\frac{\sigma}{2}\right)}{Q}^2  \leq  \average{\Delta Q_j^2}{Q} \average{ \tanh^2\left( \frac{\sigma}{2} \right)}{Q}  = \average{\Delta Q_j^2}{} \average{ \tanh^2\left( \frac{\sigma}{2} \right)}{Q} \,,
		\end{equation}
		where we used Eq.~\eqref{eq:current1} and Eq.~\eqref{eq:current2}. Now, using the chain of inequalities $ \average{ \tanh^2\left( \frac{\sigma}{2} \right)}{Q}\leq\tanh\norbra{\average{ \frac{\sigma}{2}\tanh\left( \frac{\sigma}{2} \right)}{Q}} \leq \tanh\norbra{\frac{ \average{\sigma}{}}{2}}$  (see Ref.~\cite{Hasegawa19} for a proof), we obtain from Eq.~\eqref{eq:CS}:
		\begin{equation}
			\frac{\average{\Delta Q_j^2}{}}{\average{\Delta Q_j }{}^2} \geq \norbra{\tanh\left(\frac{\langle \sigma \rangle}{2} \right)}^{-1}\,.
		\end{equation}
		Finally, defining ${\rm Var}[\Delta Q_j] = \langle \Delta Q_j^2 \rangle - \langle \Delta Q_j \rangle^2$, we obtain the TUR:
		\begin{equation}
			\frac{{\rm Var}[\Delta Q_j]}{\langle \Delta Q_j \rangle^2} = \frac{\average{\Delta Q_j^2}{}}{\average{\Delta Q_j }{}^2} -1\geq\norbra{\tanh\left(\frac{\langle \sigma \rangle}{2} \right)}^{-1}-1 = \frac{2}{e^{\langle \sigma \rangle} - 1}\,.
		\end{equation}
		This proves Eq.~\eqref{eq:FTUR} in the main text upon substituting the explicit expression of $\sigma = \sum_i \delta \lambda_i \Delta Q_i + \Delta$.
		
		\section{Allowed interaction for the generalized Gibbs state of a qubit}\label{app:int}
		Let us characterize the interactions allowed by Eq.~\eqref{eq:commutation} for two spin-1/2 particles with Hamiltonian $H = \id$, and with two non-commuting charges corresponding to $Q_1= \sigma_z$ and $Q_2=\sigma_x$. In order to do so, we use the fact that the product of Pauli matrices are a basis for the space of Hermitian operator on two qubits, so we can decompose the interaction Hamiltonian as:
			$H_{\rm int} = \sum_{i,j= 0}^3\; C_{i j} \,\sigma_i\otimes\sigma_j\,$,
		where we use the usual ordering $\sigma_0 = \id/\sqrt{2}$, $\sigma_1 = \sigma_x/\sqrt{2}$, $\sigma_2 = \sigma_y/\sqrt{2}$ and $\sigma_3 = \sigma_z/\sqrt{2}$ (the extra factor ensures orthonormalization according to the Hilbert-Schmidt inner product). A tedious but straightforward calculation shows that in order for Eq.~\eqref{eq:commutation} to hold, $H_{\rm int}$ has to be of the form:
		\begin{align} \label{eq:interaction-example}
			H_{\rm int} = C_0 \id + C_1 \sum_{i=1}^3 \sigma_i\otimes\sigma_i\,.
		\end{align}
		Then, the corresponding unitaries are of the form $U = e^{i (C_0-C_1)/2} {\rm SWAP[C_1]}$, where we introduce the generalized swap operation, given in matrix form by:
		\begin{align}
			{\rm SWAP}[\alpha] = \begin{pmatrix}
				e^{i\alpha} & 0 & 0 &0\\
				0 & \cos \alpha & i\sin\alpha&0\\
				0 & i\sin\alpha & \cos\alpha &0\\
				0 & 0 & 0 &e^{i\alpha}
			\end{pmatrix}\,.
		\end{align}
		For the simulations, we ignored the extra phase $e^{i (C_0-C_1)/2}$, as it does not contribute to any of the quantities of interest, and we set $\alpha=1$. 
		
		{
			An explicit proof that the two charges are indeed conserved can be provided by checking their commutation with the interaction Hamiltonian in Eq.~\eqref{eq:interaction-example}: 
			\begin{align}
				&[H_{\textrm{int}}, Q_1^A + Q_1^B] = C_1 [\sum_{i= x,y,z} \sigma_i^A \otimes \sigma_i^B ,\sigma_z^A + \sigma_z^B] = C_1 [\sum_{i= x,y} \sigma_i^A \otimes \sigma_i^B ,\sigma_z^A + \sigma_z^B] \nonumber \\
				&= C_1(-i \sigma_y^A\otimes \sigma_x^B - i \sigma_x^A\otimes \sigma_y^B + i \sigma_x^A\otimes \sigma_y^B + i \sigma_y^A\otimes \sigma_x^B) = 0, 
			\end{align}
			and
			\begin{align}
				&[H_{\textrm{int}}, Q_2^A + Q_2^B] = C_1 [\sum_{i= x,y,z} \sigma_i^A \otimes \sigma_i^B ,\sigma =a_x^B] = C_1 [\sum_{i= y,z} \sigma_i^A \otimes \sigma_i^B ,\sigma_x^A + \sigma _x^A + \sigma_x^B] =  \nonumber \\
				&= C_1 (-i \sigma_z^A\otimes \sigma_y^B - i \sigma_y^A\otimes \sigma_z^B + i \sigma_y^A\otimes \sigma_z^B + i \sigma_z^A\otimes \sigma_y^B) = 0.
			\end{align}

		}

\end{document}